\documentclass[conference]{IEEEtran}
\IEEEoverridecommandlockouts
\usepackage{cite}

\usepackage{lipsum} 
\usepackage{amsmath,amssymb,amsfonts}
\usepackage{algorithmic}
\usepackage{graphicx}
\usepackage{textcomp}
\usepackage{xcolor}
\usepackage{hyperref}

\def\BibTeX{{\rm B\kern-.05em{\sc i\kern-.025em b}\kern-.08em
    T\kern-.1667em\lower.7ex\hbox{E}\kern-.125emX}}
\begin{document}

\title{Lollipop: SVM Rollups on Solana\\
{\footnotesize Version: 0.1.0}
}

\author{
    \IEEEauthorblockN{
        Irvin Steve Cardenas
    }
    \IEEEauthorblockA{
        \textit{Popsicle Network
            \textsuperscript{*}
            \thanks{*This is joint research between Popsicle Network and MultiAdapative. Email: research@popsicle.network}
        }
    }
    % \IEEEauthorblockA{ contact@popsicle.network }
    \and
    
    \IEEEauthorblockN{ 
        Yugart Song
    }
    \IEEEauthorblockA{
        \textit{MultiAdaptive}
    }
}

\maketitle

\begin{abstract}
We present a formal specification for the implementation of Solana virtual machine (SVM) rollups deployed on top of the Solana Layer 1 (L1) blockchain. We further discuss our motivation, implementation, design decisions, limitations, and preliminary results. Overall, this paper is intended to serve as an initial introduction to building such system(s) on top of the Solana L1 blockchain, but does not represent an absolute. Lastly, we comment discuss on extensions of this specification to support SVM rollups on other well-established L1 blockchains systems such as Ethereum.
\end{abstract}


\section{Introduction}
For all intents and purposes, we begin by positioning our work based on the following premise --  blockchain scalability through Layer 2 (L2) solutions is either: (1) pursued by choice from the protocol developer(s), or (2) due to the inability to develop solutions that scale the base layer. The former can be observed in the case of the Bitcoin blockchain whose underlying protocol has almost reached ossification, and the pursuit towards layered or sidechain scalability has been widely accepted. One can say that this is widely accepted due to the nature of Bitcoin as the first and most reliable blockchain, whose initial recipe inspired subsequent efforts to develop blockchain technology. On the other hand, we can strongly observe that most of the industry outside of Bitcoin is focused on L2 solutions due to the inability to innovate, and sometimes reluctance to innovate, at the base layer (L1). In particular, we refer to the lack of innovation or research towards solving key bottlenecks of these distributed systems without the need for overly complex solutions.

For example, we can consider that for the most part, present blockchain systems rely on distributed consensus protocols that are mere extensions or modifications of 1980's research from pioneers such as Lamport and Liskov \cite{lamport_1982_byzantine_generals_problem, lamport_1980_reaching_agreement, liskov_1999_practical_byzantine_fault_tolerance, liskov_2002_pbft_proactive_recovery} i.e. known as classical consensus. These are closed committee protocols that rely on all-to-all communication. Examples of these are Practical Byzantine Fault Tolerance (PBFT), HotStuff and Tendermint consensus. In practice, what we presently observe in the industry are: (1) blockchain systems that leverage classical consensus and have a constrained validator set -- small enough to avoid $O(n^2)$ communication problem, or (2) blockchain systems that modify consensus parameters to portray a larger validator/committee size, but in reality simply leverage a rotating sub-committee during the voting phase. 

As far as the literature goes, the next greatest innovation since classical consensus was presented by Nakamoto \cite{nakamoto_2008_bitcoin}. He presented a consensus model that applied proof-of-work as the novel sybil resistance mechanism alongside the "longest chain" rule. Unlike classical consensus models, Nakamoto consensus allows for an open and permissionless consensus committee and provides probabilistic rather than deterministic guarnatees. However, it is worthy to note that in recent years, a novel set of consensus protocols have also been introduced, known as the the Snow Family of consensus protocols. These protocols are based on randomized sampling and metastable decision making \cite{avalanche_rocket2019}. Similarly to Nakamoto consensus, the latter consensus protocols are open and permissionless.

Nonetheless, we highlight that there's more to building a highly scalable real world systems than just theory. Although the Solana blockchain \cite{yakovenko_solana_2018} leverages a modification of classical consensus (i.e. PBFT), it introduces various key architectural innovations and mechanisms to improving block finality and transaction throughput. Pointedly, this is a stark difference from traditional Ethereum Virtual Machine (EVM) blockchain systems which have maintained the same underlying core VM architecture. Often than not, we observe that various blockchain systems require overly complex engineering to remediate performance issues not only due to their inherent architecture, but also type 1 decisions -- i.e. irreversible engineering decisions made early on in the development of the system. In the context of Ethereum, we note that the initial architect and developer openly comments that the initial version of the Ethereum blockchain was to be considered a ``technology demo'' \cite{gavin_kusamarian, wood_ethereum_2014}. This is further highlighted by Wood's later efforts to develop Polkadot -- a system that allows for a network of blockchains that are upgradeable without forks and which are supported by a shared security layer \cite{wood_polkadot_2016}. In his most recent paper, Wood discusses the constraints of Ethereum, the EVM and Polkadot \cite{wood_join_accumulate_2024}. We leave it to the reader to further explore his paper and this topic.

Besides it's vibrant growing community and ecosystem, Solana presents a different take on the VM and leverages well-established software engineering paradigms and technologies. In particular, its smart contract programming model is easy to grasp and scale. For example, it applies the well-established design principle of ``separation of concerns'' by decoupling smart contract state and logic. Furthermore, core services, standards, and applications related to payments and wallet integration are focused on user experience (UX) and simplicity. For example, one can say that the decision to create on-chain token standards allowed for services such as Candy Machine \cite{candy_machine} to be developed. The latter made it easy for anyone to deploy non-fungible tokens (NFTs) with just a steps and without any programming knowledge. This is contrary to the EVM model, where token standards are based on social consensus and require the deployment of smart contracts for every new tokens. We further cover Solana in section \ref{solana}.

In light of this, we do not present this work because Solana needs to ``scale'' or because innovation at the Solana base layer has stagnated. Rather, we present this work because there's an opportunity for applications to leverage the Solana development stack and develop SVM applications in isolation, while still benefiting from the economic security and consensus of the Solana L1 blockchain, all while contributing to the Solana ecosystem. This paper presents a formal specification of rollups on Solana, and makes a best effort to present the reader with enough literature and context to understand how to build such rollups. Our hope is that this contribution can further expand the adoption of the Solana virtual machine, technology stack and grow the Solana ecosystem.

The paper is structured as follows. In section \ref{background}, we review the literature required to understand our system. In section \ref{network_model}, we present our architecture and threat model. We later formalize our fraud proof system in section \ref{fraud-proof} and discuss data availability in section \ref{data-availability}. Finally, we end by reviewing our ongoing work in section \ref{discussion_future_work}

\section{Background and Motivation} \label{background}

\subsection{Solana}\label{solana}
The Solana blockchain \cite{yakovenko_solana_2018} presents a novel blockchain architecture in comparison to previous systems designed around the Ethereum virtual machine (EVM). Solana implements proof-of-stake (PoS) as a sybil control mechanism, alongside one of it's key innovations -- the proof-of-history (PoH) algorithm. PoH is a type of verifiable delay function (VDF) that allows for the ordering and timestamping of transactions sent over the network. Beyond the latter, Solana stands apart for it's use of high-performance hardware, mempool-less transaction forwarding protocol (Gulf Stream) and its different take on the traditional blockchain account model -- which resembles the Linux operating system file system.

Taking the file system model into consideration, all accounts can be viewed as files stored in Solana's validators. These files either store information, or an executable binary. Respectively, these files represent two different types of accounts: (1) non-executable and (2) executable accounts. These files have metadata associated which describes ownership details, rent epoch, and executable status. Given that these files take up memory, they pay rent in Solana's native token SOL.

This account model allows for decoupling of smart contract logic and state -- i.e. stateless smart contracts. Whereby, developers only focus on writing smart contract logic that interacts with references to accounts passed as inputs. This is similar to traditional software engineering where business logic is agnostic and separated from the database where the data is stored. But, this is contrary to the traditional EVM model where state and smart contract logic are coupled.

For brevity, we highlight that Solana’s ability to achieve high throughput low latency is in part due to its stringent hardware requirements. Although initial criticism surrounded these requirements, it’s worthy to note that recent Ethereum Layer 2 solutions have began proposing a similar concept - “beefy" nodes or sequencers, that leverage high performance hardware. Further details on Solana and it's documentation can be found here \cite{yakovenko_solana_2018, solana_docs}

\subsection{Merkle Trees and Sparse Merkle Trees} \label{sparse-merkle-tree}
A sparse merkle tree (SMT) \cite{laurie_kasper_2012} is a type of cryptographic data structure that combines both a traditional Merkle tree  and a Patricia tree data structure to efficiently store a large set of key-value pairs. The advantage of a SMT over a traditional Merkle tree is that tree nodes are organized in a way that only certain nodes need to be stored, typically those that contain non-empty values or that are on the path to a leaf node with a value — this is what defines them as “sparse”. Overall, they allow for compact proofs (Merkle proofs) that a particular key-value pair is in the tree or not, without revealing the values themselves. This leads to the optimization of storage and computational efficiency.  The relevant use cases for SMTs in Layer 2 solutions are for state commitment and state verification. State commitments refers to Layer 2 solutions committing the state of off-chain transactions (off the Layer 1). By committing to this state on the Layer 1, participants of the Layer 2 can prove the validity of their off-chain transactions without revealing the entire state or requiring the Layer 1 blockchain to process every off-chain transaction.

\subsection{KZG Polynomial Commitment}
Polynomials are mathematical expressions that consist of a set of variables and coefficients. In the context of blockchain technology, polynomials can be used to indirectly represent various types of data, such as account balances or transaction details, through their coefficients. Such representations enables secure manipulation of committed data using polynomial operations. This preserves the privacy of the underlying information while ensuring its integrity. Polynomials serve as the foundation for KZG commitments \cite{kzg_2010} discussed below.

A KZG commitment, short for Kate-Zaverucha-Goldberg commitments, is a cryptographic commitment scheme that leverages polynomial interpolation and evaluation techniques. In essence, KZG commitments allow for the representation of data through polynomial coefficients, facilitating secure and efficient operations on committed data without revealing the original values. This is achieved by generating commitments through polynomial evaluation at specific points, resulting in single values that compactly represent the underlying data. These means that such commitments can represent large amounts of data efficiently and compactly.

\subsection{SVM Rollups on solana}
Solana is one of the few blockchain networks that is high-performant in isolation and as a public network. In Particular we highlight the rise of application-specific SVM blockchain networks such as Pyth Network \cite{solana_pyth_network}, and Cube Exchange \cite{solana_cube_exchange}. Within the Solana ecosystem, the latter two networks are defined as Solana Permissioned Environments (SPEs). In short, these are specialized private environments tailored to the particular needs of an application. Overall, SPEs allow for customization and control. Similarly, we can implement rollup networks that provide the same level of customization, control and flexibility, but which rely on Solana's economic security. Contrary to traditional rollup stacks, in \cite{magnet_sdk} we present a novel framework that allows for elastic and on-demand block generation. This increases cost-efficiency and scalability and deems such rollups ``smart'', given the fact that rollups develop with our framework can adjust its parameters based on the overall state of the L1 and L2 -- e.g. transaction costs and network activity, respectively.

\begin{figure}[htbp]
\centerline{\includegraphics[width=8cm,height=4cm]{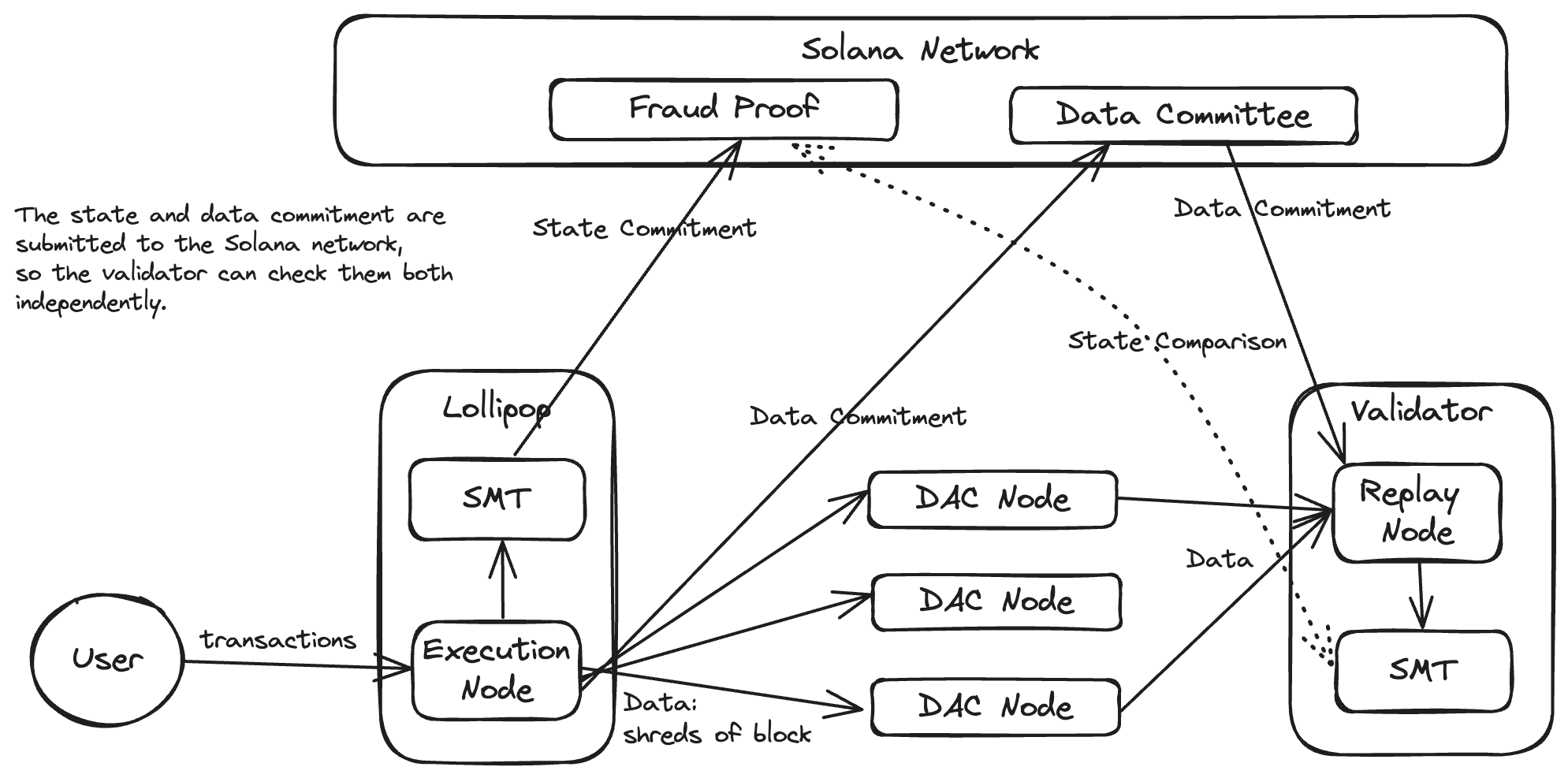}}
\caption{Architecture of Lollipop}
\label{fig}
\end{figure}

\section{Network Model} \label{network_model}
There are two classes of nodes in the Lollipop rollup system: execution nodes, and validator nodes.

\begin{enumerate}
    \item Execution Nodes: These nodes commit the state of the rollup to a Solana smart contract according to slot and also commit the transaction data processed during this period to Solana. Execution Nodes can be seen as a type of sequencer node within the Ethereum literature.

    \item Validator Nodes: These nodes first retrieve the transaction data from the data availability service (e.g. DAC or a third-party DA) according to the commitment of the transaction data and then execute these transactions. If a validator node finds that its local state does not match the commitment submitted by the execution node, it will apply for a fraud-proof challenge in the Solana contract to determine whether the execution node is misbehaving.
\end{enumerate}

\subsection{Threat Model}
We make the following assumptions in our threat model:
\begin{itemize}
\item Solana Network: The Solana network is the only component we trust. Lollipop submits the processed transaction data and execution results (as commitments) to Solana. The transaction data is directly trusted, and the Lollipop network uses this data as the sole basis for consensus. The execution node and the validator node need to agree on the execution result on Solana so that the final settlement can be completed. If no consensus can be reached, it is up to the Solana network to determine who is lying.
    
\item Execution Node: Execution nodes can also be untrusted. This is node is responsible for the collection, packaging, and execution of transactions, but we cannot guarantee that it is completely trustworthy. Thus, we verify the execution results submitted by it to ensure the safety of users' funds.

\item Validator Node: Validator nodes can submit challenges. There are multiple validator nodes, and we assume that at least one of them is honest. When a validator node challenges the execution result, the execution node needs to prove that the submitted result is correct. In order to prevent the validator node from maliciously submitting challenges and consuming the resources of the execution node, we require the validator node to stake tokens before launching the challenge. If the execution node can prove that it has not cheated, the tokens of the validator node will be slashed.
\end{itemize}

\section{Fraud Proof} \label{fraud-proof}
Most blockchains like Ethereum organize state data into tree data structures (e.g. Merkle trees, Verkle trees, etc). This data includes accounts, balances, and smart contracts. In the context of traditional rollups, the world state is represented by the Merkle root. On the contrary, the Solana blockchain does not maintain a committed global state tree.

The good news is that the Solana blockchain outputs the required account information that needs to be modified after processing each transaction. We define this information as $ma$. We can therefore use this information to build a sparse Merkle tree (SMT). Let's assume that the root of the SMT is $r$.

\subsection{State Commitment}
We therefore formulate a state commitment as follows. At slot $i$, we first get the root $r_i$ of the SMT after executing all the transactions in the current slot. We then calculate the hash of transactions in the current slot, as defined by $HA_i$. If there are $t$ transactions in the slot, then we can say that: $t_{i,1},...,t_{i,t}$, so the $HA_i=H_{i,t}$ where $H_{i,j}=H(tx_{i,j}+ma_{i,j}+H_{i,j-1})$, and $H_{i,0}=NULL$, where $H$ is an arbitrary hash function. Finally, we submit $\{r_i,HA_i,t_i\}$ to our Solana contract as the state commitment.

\subsection{Interactive Proof}
When a validator node applies for a fraud-proof challenge at slot $i$, it must be that the validator and execution nodes agree with the state root $r_{i-1}$, but disagree with $r_i$. Since there are $t$ transactions in slot $i$, we must use an interactive proof to find the conflicting transaction between these two nodes.

Using dichotomy, the execution node continuously submits the intermediate state commitment $\{r_{i,j}, H_{i,j}\}$ to ask the validator node to choose to agree or disagree. $r_{i,j}$ is the root for the SMT after executing $j$ transactions in slot $i$. At last, we can get $j$, so the two nodes agree on $\{r_{i,j}, H_{i,j}\}$, but disagree on $\{r_{i,j+1}, H_{i,j+1}\}$. So the conflicting transaction is $t_{i,j}$.

\subsection{Replay a Transaction}
Since the conflicting transaction is found, now we need to replay this transaction to determine which node is lying.

Before that, we note that we had already compiled a SVM into rBPF (Solana bytecote) \cite{solana_rbpf} as part of our Solana contract.

First, the execution node will get the input of the transaction $t_{i,j+1}$, where the input is all the accounts the transaction will read or write. Then it will get the Merkle proof of the input based on the SMT with root $r_{i,j}$.

Second, the execution node will submit the input and input proof to the Solana contract. 

The contract first checks the input by input proof, then executes the transaction $t_{i,j+1}$ with the input and gets the output, which is $ma_{i,j+1}$.

So the contract can check if $H_{i,j+1}==H(tx_{i,j+1}+ma_{i,j+1}+H_{i,j})$. If so, it continues to check that if $r_{i,j} + ma_{i,j+1} ==> r_{i,j+1}$. In fact, the key of $ma_{i,j+1}$ is in the list of input, but only the value is modified, so the Merkle proof of input is also the proof of $ma_{i,j+1}$ based on the SMT with new root. 

\begin{figure}[htbp]
\centerline{\includegraphics[width=8cm,height=4cm]{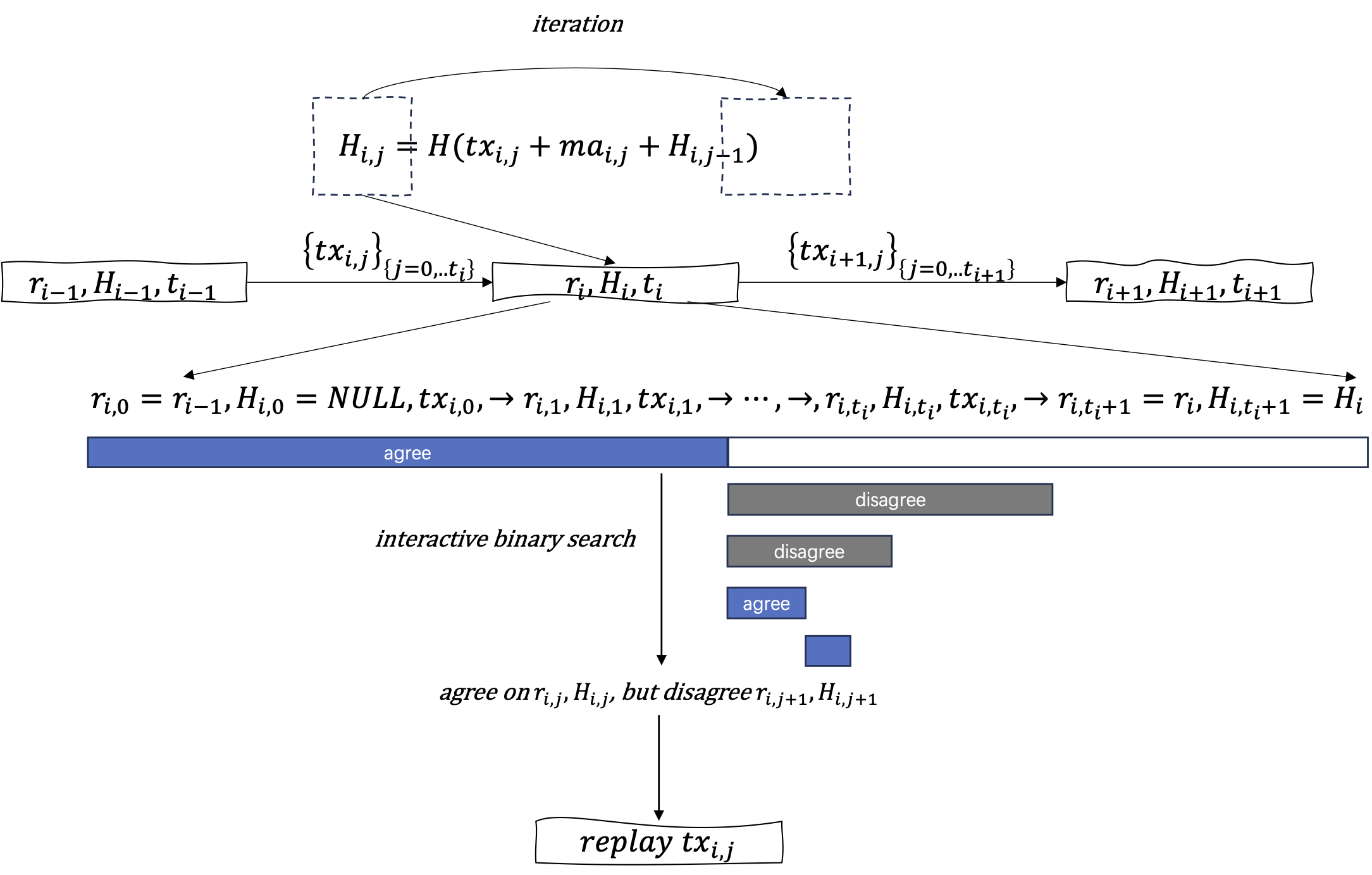}}
\caption{Fraud Proof}
\label{fig}
\end{figure}

The execution node can be considered honest only if it passes all the checks of the contract, otherwise it is lying.

\section{Data Availability} \label{data-availability}
Solana has a strict limit on the size of transactions of 1232 bytes \cite{solana_transactions_documentation}, which does not meet the needs for batch submission of transaction data from Lollipop. Therefore, Lollipop only submits the commitment of data to Solana, and stores the data in a data availability committee (DAC).

\subsection{Data Commitment}
A DAC is a collection of permissioned entities tasked with holding copies of the blockchain data offline.
 
Lollipop first encodes the data to a polynomial, and generates the KZG commitment of it. Lollipop then sends the data and commitment to the members of DAC and requests for signatures. Once the data and commitment are placed, member of the DAC will first check that they are a pair -- check the the commitment is from the corresponding data; then sign the commitment back to Lollipop. Lollipop will check and collect the signatures, and finally submit the commitment and signatures to the Solana contract.

\subsection{Data Audit}
If we want to check whether the members of the DAC have stored all the data correctly, we can use the following data audit method.

The Solana contract records the commitments $cm_i$ corresponding to each data submission and the corresponding polynomials $f_i(x)$. Assuming data has been submitted a total of $t$ times -- when the project needs to initiate a full audit, they must generate $t$ random numbers $r_i$ and a random value $v$. The storage provider is then required to open the polynomial $F(x) = \sum r_i \cdot f_i(x)$ at the value $v$.

The commitment corresponding to $F(x)$ is $CM = \sum r_i \cdot cm_i$. When the value of $t$ is large, it can be challenging to calculate this commitment within the contract. Therefore, an interactive challenge protocol is designed to ensure that the project and the storage provider can reach consensus on $CM$.

First, the project submits $CM_{0,t} = \sum_{i=0}^{t} r_i \cdot cm_i$. If the storage provider disagrees, the project submits $CM_{0,t/2}$ in an interactive binary manner, allowing both parties to find a pair ${CM_{i,j}, CM_{i,j+1}}$ that they agree upon $CM_{i,j}$ but not on $CM_{i,j+1}$.

At this point, the contract only needs to verify whether $CM_{i,j+1} = CM_{i,j} + r_{j+1} \cdot cm_{j+1}$. This determination can reveal who is being dishonest, leading to penalties.

Once consensus is reached, the storage node must generate $F(v)$ and its corresponding proof for contract verification.

\subsection{Data Sampling}
To avoid network connection failure caused by retrieving large amounts of data we adopt a data sampling scheme. We randomly request only parts of the data each time, and finally synthesize the complete data.

Because each data is encoded into a polynomial $f_i(x)$, and their degree is at most $n$, you only need to know the values of $f_i(x)$ at up to $n+1$ points to reconstruct $f_i(x)$.

So when sampling data, for each data, the sampling party randomly generates up to $(n+1)/16$ values $v_j$ and then requests broadcast nodes to open them at these $(n+1)/16$ points. They provide a commitment Merkle proof, 

$$
[cm_i, [{v_j}, \pi], proof]
$$

for that piece of data. The sampling party first verifies the legality of $cm_i$ based on the Merkle proof and then performs KZG verification.

Because the random space has a size of $2^{256}$, we can assume that the sampling party will not randomly choose the same value twice. Therefore, with just 16 sampling parties, you can obtain the values of $f_i(x)$ at $n+1$ points, ensuring the reconstruction of $f_i(x)$.

\section{Discussion \& Future Work} \label{discussion_future_work}

\subsection{Hardware Requirements}
Compared to blockchains such as Bitcoin and Ethereum, Solana achieves higher transaction per second (TPS), but requires more costly servers and even GPUs.

Based on Solana's official recommendations, Lollipop's execution nodes will use 16-core CPUs, 512G memory, 1TB pd-SSD disks, and 2x Nvidia V100 GPUs.

Validator nodes do not participate in consensus and do not need to process votes. They only need to verify the results committed to the Solana network. To reduce the workload of the validator nodes, we can even let them use a sampling check mode to check the honesty of the execution nodes.

The execution node can submit the transaction data and the state diff during this period to the DA. After the validator node reads the state diff from the DA, it uses it to update its local state tree, and then randomly replays the transactions in a slot to see if the results are consistent. This can reduce the burden on the validator node and allow the verification node to run on a lower cost server.

\subsection{Compatability}
With the development of zero-knoweldge proof (ZKP) technology, many zkVM projects have emerged, such as Starknet \cite{starknet_website}, zkSync \cite{zksync_website}, Polygon \cite{polygon_zkevm}, Scroll \cite{scroll_zkevm}, which can generate ZK proofs for the execution process of their VMs (cario-VM and EVM). Subsequently, ZK development tools for languages, such as RISC0 \cite{risc_zero} and SP1 \cite{sp1}, have also appeared. Therefore, we will use SP1 to generate proofs for the replay process of the last part of our fraud proof, so that verification can be completed on other L1s that support smart contracts. In other words, our solution can easily migrate Lollipop to other L1 ecosystems with the help of ZKP technology.

\subsection{Data Availability}
Ongoing work focuses on exploring third-party data availability solutions such as Celestia \cite{celestia_site}, NEAR DA \cite{near_da}, and MultiAdaptive DA \cite{domicon_whitepaper}.

In principle, we do not want to introduce new security assumptions. We hope that the security of Lollipop can be directly built on the consensus of the Solana mainnet. Therefore, our requirement for third-party DA is that it can achieve the following:

1. Confirm which data should be stored in the DA on the Solana network.

2. Slash DA nodes that do not perform storage and broadcast tasks on the Solana network.

Unfortunately, neither Celestia nor NEAR DA can provide this security measurements. They complete the above operations on their own blockchains, and then notify other L1s through a bridge, which data they store, and slash dishonest nodes on their own blockchains.

On the other hand, we found that MultiAdaptive DA provides a reliable, native secure solution. It can realize all our needs in the Solana contract without bringing additional security assumptions to Lollipop. Furthermore, they also provide a 100\char\% data audit mechanism to ensure that historical data is fully preserved.

\subsection{Decentralized Shared Sequencers}

Popsicle Network's Decentralized Shared Sequencer (DSS) is a decentralized infrastructure for rollups based on shared security. Anyone within the Popsicle ecosystem can join the network as a sequencer candidate without permission, and the final sequencers are elected through public elections based on the holders of the base chain. The network then groups the elected sequencers according to the predetermined number of Sequencer Groups using a verifiable random function (VRF) algorithm. Ultimately, Popsicle shares sequencers in terms of groups to rollups in the Popsicle ecosystem, providing sequencing services for projects leveraging this infrastructure.

By allowing permissionless registration, the risk of concentrated power is diluted, giving everyone an equal opportunity to become a sequencer candidate. In the case of Solana as a base blockchain infrastructure, staking elections allow for stronger security guarantees inherited from the Solana L1 blockchain. Additional, incentive mechanisms are used to elect the highest-quality sequencer service providers. Finally, by providing services in the form of groups, the platform enhances its resistance against centralization. In this way, Popsicle Network and it's DSS offers the best anti-censorship and decentralization solutions for an ecosystem of rollups, while further reducing the difficulty of deploying and operating rollups. This DSS infrastructure can be applied to different base blockchains such as Solana -- our ongoing work particularly focuses on the Bitcoin blockchain. \cite{popsicle_net_sequencers}

\subsection{Solana VM Bitcoin Layer 2}
Ongoing work at Popsicle Network focuses on applying similar principles and design to develop Bitcoin Layer 2 solutions that are more (1) trustless, (2) permissionless and (3) decentralized. In this context, our ongoing work focuses on developing SVM Layer 2 networks on top of Bitcoin. Various complexities arise given the restrictions around Bitcoin Script and executable logic, nonetheless Popsicle Network explores this and presents viable solutions in a separate paper.

% \section*{Acknowledgment}
% Thanks *** for help .....

\bibliographystyle{IEEEtran}
\bibliography{ref}

\end{document}